# A mathematical model integrates diverging PXY and MP interactions in cambium development

Kristine S. Bagdassarian[1], J. Peter Etchells[1], and Natasha S. Savage*[2]

*Corresponding author
[1] Durham University, Department of Biosciences, Durham, UK
[2] University of Liverpool, Liverpool, UK

## Abstract

The cambium is a meristematic tissue in plant stems. Here, cell divisions occur that are required for radial growth of plant stems. Daughters of cell divisions within the cambium differentiate into woody xylem cells towards the inside of the stem, or phloem towards the outside. As such, a pattern of xylem-cambium-phloem is present along the radial axis of the stem. A ligand-receptor pair, TDIF-PXY promotes cell division in the cambium, as do the phytohormones, cytokinin and auxin. An auxin response factor, MP, has been proposed to initiate cambial cell divisions by promoting PXY expression, however, MP has also been reported to repress cambial cell divisions later in development where TDIF-PXY complexes are also reported to suppress MP activity. Here, we used a mathematical modelling approach to investigate how MP cell division-promoting activity and cell division-repressing activity might be integrated into the same network as a negative feedback loop. In our model, this feedback loop improved the ability of the cambium to pattern correctly and was found to be required for normal patterning when MP was stable. The implications of this model in early and late cambium development are discussed.

## 1. Introduction

Most terrestrial plant biomass is derived from the cambium, which promotes radial, secondary growth of stems in gymnosperms and woody angiosperms. The cambium constitutes a bifacial meristem from which phloem and xylem tissues are derived. As stem cells in the cambium divide, their daughters differentiate into phloem towards the outside of the stem, or xylem towards the inside (Esau, 1960; Esau, 1965; Evert, 2006). The cambium represents an atypical stem cell population by virtue of its bifacial nature, but also because it arises post embryonically, forming de novo following germination (Baum et al., 2002). As such, cambial stem cells are defined within a pre-patterned tissue that arises during embryogenesis (De Rybel et al., 2014; Scheres et al., 1994). In Arabidopsis roots the pre-pattern constitutes a central file of xylem cells with phloem poles on either side (Fig 1A). Each phloem pole is separated from the central xylem by a row of procambium cells (Dolan et al., 1993). Formation of the cambium at initiation of secondary growth occurs when the cells of xylem identity act as "organiser cells", promoting their neighbours to form the cambial meristem and divide (Smetana et al., 2019). This marks the transition from primary to secondary growth.

The events that promote the transition from primary to secondary growth start with the formation of an auxin maximum in the organizer cells which have xylem identity (Smetana et al., 2019). Auxin responses are mediated by transcription factors in the Auxin Response Factor (ARF) family (Roosjen et al., 2018), and the auxin maximum in the organiser cells results in activation of MONOPTEROS (MP; also known as ARF5)(Smetana et al., 2019). MP promotes expression of auxin efflux carriers, members of the PIN family (Bhatia et al., 2016; Brackmann et al., 2018; Hardtke and Berleth, 1998; Przemeck et al., 1996). Thus, auxin promotes its own movement via an auxin, MP, PIN pathway. Auxin, acting through MP, also activates expression of members of the HD-Zip-III family of transcription factors and PHLOEM INTERCALATED WITH XYLEM (PXY) receptor kinase (Baima et al., 1995; Carlsbecker et al., 2010; Donner et al., 2009; Izhaki and Bowman, 2007; Mattsson et al., 2003; Ohashi-Ito and Fukuda, 2003; Smetana et al., 2019; Ursache et al., 2014; Zhou et al., 2007). PXY expression is promoted by HD-Zip-III transcription factors (Smetana et al., 2019), thus MP sits at the top of a feed-forward loop in which MP activates HD-Zip-III's, and both MP and HD-Zip-III's promote PXY expression. Activation of PXY expression promotes the first cell divisions of the cambium (Figure 1E) (Smetana et al., 2019).

PXY is a central regulator of vascular proliferation in plant vascular tissue (Etchells and Turner, 2010; Hirakawa et al., 2008). Its cognate ligand, Trans-Differentiation Inhibitory Factor (TDIF), is a dodecapeptide derived from two genes, CLE41 and CLE44 that are expressed in the phloem (Ito et al., 2006). TDIF-PXY complexes are in an active state and promote cell divisions, both by excluding xylem differentiation from the cambium stem cells, and by promoting the divisions themselves (Etchells and Turner, 2010; Hirakawa et al., 2008). BES1, a transcription factor that promotes xylem activity is degraded upon TDIF-PXY binding, thus preventing premature xylem differentiation (Kondo et al., 2014). TDIF-PXY signalling promotes cell division by activating expression of homeodomain transcription factors WOX4 and WOX14 (Etchells et al., 2013; Hirakawa et al., 2010; Smit et al., 2020).

Cytokinin is another hormone that plays an important role in PIN regulation (Bishopp et al., 2011a; Marhavý et al., 2014; Pernisová et al., 2009; Růžička et al., 2009; Šimášková et al., 2015), cambium formation, and subsequent cambium cell divisions (Hejátko et al., 2009; Matsumoto-Kitano et al., 2008; Ye et al., 2021). Indeed, mutants lacking cytokinin biosynthesis never form a cambium (Matsumoto-Kitano et al., 2008). Antagonistic relationships between auxin and cytokinin feature in pattern formation across plant vascular tissue (Schaller et al., 2015). Auxin signalling induces ARABIDOPSIS HISTIDINE PHOSPHOTRANSFER PROTEIN 6 (AHP6), a pseudophosphotransfer protein which acts to dampen cytokinin signalling (Mähönen et al., 2006; Moreira et al., 2013; Suzuki et al., 1998; Werner et al., 2006). Thus, in cells with an auxin maximum, such as those of xylem identity during primary growth, cytokinin signalling is attenuated (Mähönen et al., 2006; Matsumoto-Kitano et al., 2008; Moreira et al., 2013; Suzuki et al., 1998). In turn, cytokinin suppresses auxin active transport through decreasing the PIN levels at the post-transcriptional stage (Bishopp et al., 2011b; Ioio et al., 2008; Marhavý et al., 2011; Pernisová et al., 2009; Šimášková et al., 2015; Zhang et al., 2011). Cytokinin is bulk-transported down the phloem (Bishopp et al., 2011b; Hirose et al., 2008), reinforcing the pattern of maximal cytokinin signalling in the phloem, and minimal in the xylem (Immanen et al., 2016; Smetana et al., 2019; Tuominen et al., 1997; Uggla et al., 1998; Uggla et al., 1996). By contrast, once secondary growth is established, the auxin maximum is present in the cambium cell adjacent to the xylem (Immanen et al., 2016; Smetana et al., 2019; Tuominen et al., 1997; Uggla et al., 1998; Uggla et al., 1996) (Figure 1D), the same tissue in which PXY is expressed (Fisher and Turner, 2007).

As discussed above, during the transition from primary to secondary growth MP is at the top of a feed-forward loop which promotes cambial cell divisions (Figure 1A). By contrast MP has been reported to act as a repressor of cambium cell division in established vascular tissue (Figure 1B). *mp* mutants demonstrate increases in the size of cambium-derived tissues, and high MP levels were found to reduce cambial-derived tissue (Brackmann et al., 2018). Thus, the data suggests that the influence of MP on cambial cell divisions changes during development: MP promotes cambium cell division in initiating vascular tissue ((Smetana et al., 2019), Figure 1A), but MP represses cambium cell divisions in established vascular tissue ((Brackmann et al., 2018), Figure 1B). During the later stages of development, TDIF-PXY was shown to represses MP activity by preventing phosphorylation at a site that contributes to MP's activation (Han et al., 2018) (Figure 1B).

When considering the data in the form of network diagrams (Figures 1A and 1B) these seemingly contradictory data (MP acting as both promoter and repressor of cambial divisions) are not contradictory at all. Rather, they support the hypothesis whereby MP directly represses cambial cell divisions (Brackmann et al., 2018) and MP promotes the activation of PXY, active PXY then promotes cambial cell divisions (Smetana et al., 2019) via the prevention of MP activation (Han et al., 2018). As such these interactions constitute an MP negative feedback loop, whereby MP promotes its own down regulation; MP promotion of PXY, and PXY suppression of MP (Figure 1C). In this paper the consequence of including these regulatory interactions in the same network was tested using mathematical modelling. The model was built using the cellular organisation present in the initiating cambium, just after the transition into secondary growth. It was posited that if the MP feedback loop was indeed present in this tissue, then its presence would improve the ability of the cambium to maintain cytokinin and auxin concentration profiles (Figure 1D) (Fu et al., 2021; Immanen et al., 2016; Smetana et al., 2019; Tuominen et al., 1997; Uggla et al., 1998; Uggla et al., 1996). Including MP negative feedback did improve the ability of the modelled tissue to pattern correctly and furthermore, the addition of MP negative feedback never hindered the modelled tissues ability to pattern correctly. Interestingly, when MP was stable, MP negative feedback was required for the modelled tissue to pattern correctly. One explanation of these findings is that the MP-PXY-MP feedback loop is present both in initiating and established cambium. Other factors, such as differences in tissue topology, or molecular actors, may then influence the feedback loop to favour MP activation of PXY early in development, but PXY repression of MP in established tissue, will be a matter for future study.

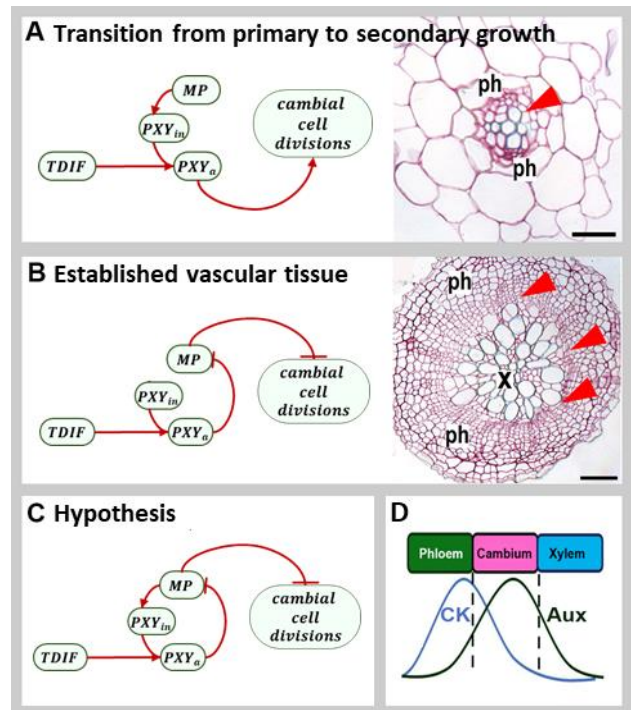

**Figure 1. Diagrams showing interactions between PXY and MP in secondary growth, and tissue morphology in which these interactions occur.**
**(A)** MP activates PXY expression (*PXY$_{in}$*), and PXY promotes initial cambial cell divisions upon interaction with TDIF (*PXY$_a$*). Image on left shows root with initial cambial cell division. **(B)** When secondary growth is established, MP represses cambial cell division and is repressed by TDIF-PXY (*PXY$_a$*). Image shows root with an established cambium. **(C)** MP negative feedback loop. Combination of network diagrams in (A) and (B). TDIF-PXY (*PXY$_a$*) promotes cambial cell division via the prevention of MP activation. **(D)** Schematic showing an auxin (Aux) maxima on the xylem side of the cambium and cytokinin (CK) concentration highest in the phloem (adapted from Fischer et al., 2019). Scales are 20 μm (A), or 50 μm (B). X is xylem, ph is phloem, red arrowheads point to dividing cambial cells.

## 2. Materials and Methods

### 2.1 Model formulation

A one-dimensional reaction diffusion model was built to investigate the consequence of MP negative feedback on vascular tissue patterning. The model contained three, well mixed, spatial domains; the phloem, cambium and xylem. The lengths of the domains were set to; $2.6 \mu m$ for the phloem, $1.2 \mu m$ for the cambium and $10.8 \mu m$ for the xylem, which were derived from previously described *in planta* measurements (Wang et al., 2019). The model contained components which have been shown to impact vascular tissue patterning. Namely; the hormones cytokinin and auxin, the PIN proteins, the activator of PIN proteins MP, the receptor-kinase PXY and its ligand TDIF.

Both hormones, cytokinin and auxin, move between the modelled spatial domains. Cytokinin was the only component of the model which moved between the domains via diffusion. Cytokinin is rapidly transported down the phloem from the shoot to the root (Hirose et al., 2008) and is thus modelled as a constant cytokinin source. Cytokinin then diffuses from the phloem into the cambium and xylem. The diffusion coefficient of cytokinin was set to $220\ \mu m^2/s$ (Moore et al., 2015). The phloem was modelled to be a constant source of auxin as that too is transported down the phloem (Adamowski and Friml, 2015; Blakeslee et al., 2005; Blilou et al., 2005; Friml et al., 2002a; Friml et al., 2002b; Goldsmith, 1977; Kepinski and Leyser, 2005; Ljung et al., 2005; Michniewicz et al., 2007; Swarup et al., 2001; Vieten et al., 2005; Zhou and Luo, 2018). However, as auxin moves between cells via PINs, auxin movement was described as a reaction in the model. Any hormone moving from the cambium back into the phloem was considered to be transported out via bulk transport (Adamowski and Friml, 2015; Bishopp et al., 2011b; Hirose et al., 2008; Michniewicz et al., 2007), leaving the phloem hormone concentration constant. Any hormone moving from the xylem on the opposing side to the cambium, was removed from the model.

PXY's peptide ligand, TDIF, is cleaved from CLE41 and CLE44 peptides which are transcribed and translated in the phloem. Following excretion from phloem cells, TDIF moves to the cambium (Etchells and Turner, 2010; Hirakawa et al., 2008; Ito et al., 2006). As TDIF is continuously produced in the phloem, the concentration of TDIF is modelled as a constant (Etchells and Turner, 2010; Hirakawa et al., 2008; Ito et al., 2006).

Reactions included in the model were; the suppression of cytokinin by auxin (Bishopp et al., 2011a; Mähönen et al., 2006; Moreira et al., 2013; Müller and Sheen, 2008; Nordström et al., 2004; Werner et al., 2006), the release of MP inhibition by auxin (Chen et al., 2015) , the suppression of PINs by cytokinin (Bishopp et al., 2011b; Ioio et al., 2008; Müller and Sheen, 2008; Pernisová et al., 2009; Růžička et al., 2009; Šimášková et al., 2015), the activation of PINs by MP (Bhatia et al., 2016; Hardtke and Berleth, 1998; Krogan et al., 2016; Przemeck et al., 1996), the induction of PXY transcription by MP, the activation of PXY by TDIF (Etchells and Turner, 2010; Fisher and Turner, 2007; Hirakawa et al., 2008; Ito et al., 2006), and the repression of MP by activated PXY (Han et al., 2018). Transcription and translation were not separated in the model and modelled as one reaction. Reactions were modelled using mass action kinetics. The reactions included in the model are illustrated in Figure 2A and summarised in Table 1.

The concentration of cytokinin in the phloem, cambium and xylem are denoted, $[CK_p]$, $[CK_c]$ and $[CK_x]$, respectively. When denoting the diffusion of cytokinin the cytokinin concentrations are referred to as $[CK]$. $D_{CK}$ denotes the diffusion coefficient of cytokinin. $[AUX_p]$, $[AUX_c]$ and $[AUX_x]$ denote the concentrations of auxin in the phloem, cambium and xylem. The concentration of PINs and MP in the cambium and xylem are denoted, $[PIN_c]$, $[MP_c]$ and $[PIN_x]$, $[MP_x]$, respectively. The concentration of TDIF in the phloem is denoted, $[TDIF_p]$. The concentration of inactive (not bound to TDIF) and active PXY (TDIF-PXY) in the cambium are denoted, $[PXY_{in}]$ and $[PXY_a]$, respectively. $d_*$ denotes the rate of basal degradation of the component $*$. $r_i$ denotes the reaction rate of reaction $i$. Equations (1) to (10) describe the model.

In order to investigate the hypothesis that MP negative feedback could improve vascular tissue patterning, the model was solved and analysed, with and without MP negative feedback. The model with MP negative feedback was described by equations (1) to (10) with all parameter values greater than zero. The model without MP negative feedback was also described by equations (1) to (10) but with parameters which represent MP negative feedback set to zero, namely $[TDIF_p], r_7, r_8, r_9, d_{PXY_{in}}, d_{PXY_a}$ (Figure 2A, Table 1).

Numerical solutions were calculated using Euler's method on an irregular grid (see Supplement for the diffusion matrix). All numerical solutions presented in the Results section, and all codes used to solve and analyse the numerical solutions, can be found on GitHub (Bagdassarian, 2021). Closed form, steady state analysis can be found in the Results section and the Supplement.

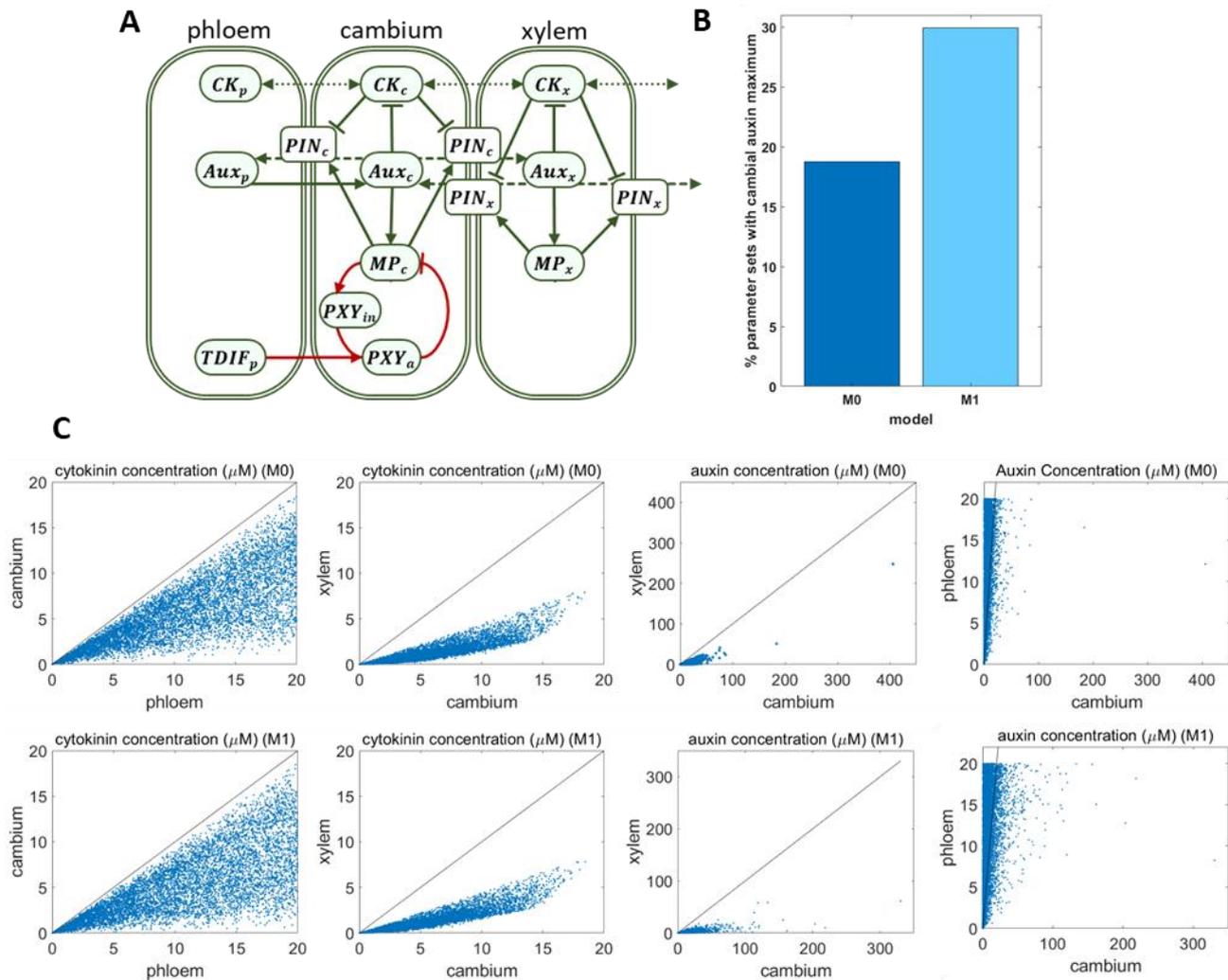

**Figure 2. (A)** Diagram showing modelled interactions. Links represent interactions. Arrows represent promotion, barbed ended links represent repression. Red links highlight the hypothesised MP negative feedback loop (Figure 1C). **(B)** Percentage of parameter sets that achieve cambium auxin maximum in the presence (M1) or absence (M0) of MP negative feedback. **(C)** Concentration comparisons of the hormones cytokinin and auxin within different tissues. 7500 data points in each graph. Each data point is the steady state solution for an individual parameter set. Black line represents the boundary when concentrations are equal.

| new reaction number | reaction units | reaction description | reference |
|---|---|---|---|
| $r_1$ | $s^{-1}$ | auxin flux from the phloem to the cambium | (Blakeslee et al., 2005; Blilou et al., 2005; Friml et al., 2002a; Friml et al., 2002b; Swarup et al., 2001) |
| $r_2$ | $\mu^{-1}s^{-1}$ | auxin transport via PINs | (Blakeslee et al., 2005; Blilou et al., 2005; Goldsmith, 1977; Vieten et al., 2005) |
| $r_3$ | $\mu^{-1}s^{-1}$ | auxin suppression of cytokinin | (Bishopp et al., 2011a; Mähönen et al., 2006; Moreira et al., 2013; Müller and Sheen, 2008; Nordström et al., 2004; Werner et al., 2006) |
| $r_4$ | $s^{-1}$ | Auxin-induced release of MP | (Chen et al., 2015; Dharmasiri et al., 2005; Peer, 2013; Ulmasov et al., 1997; Ulmasov et al., 1999; Weijers and Wagner, 2016) |
| $r_5$ | $\mu^{-1}s^{-1}$ | cytokinin suppression of pins | (Bishopp et al., 2011b; Ioio et al., 2008; Müller and Sheen, 2008; Pernisová et al., 2009; Růžička et al., 2009; Šimášková et al., 2015) |
| $r_6$ | $s^{-1}$ | MP activation of PINs | (Bhatia et al., 2016; Hardtke and Berleth, 1998; Krogan et al., 2016; Przemeck et al., 1996) |
| $r_7$ | $s^{-1}$ | MP induces PXY transcription | (Smetana et al., 2019) |
| $r_8$ | $\mu^{-1}s^{-1}$ | TDIF activates PXY | (Etchells and Turner, 2010; Fisher and Turner, 2007; Hirakawa et al., 2008; Ito et al., 2006) |
| $r_9$ | $\mu^{-1}s^{-1}$ | active PXY repression of monopteros | (Han et al., 2018) |

**Table 1. Modelled reactions.** Table describing the reactions included in the model, their references, their reaction number and units.

$$\frac{d}{dt}[Aux_c] = r_1[Aux_p] + \frac{r_2}{2}[PIN_x][Aux_x] - r_2[Aux_c][PIN_c] - d_{Aux}[Aux_c] \qquad (1)$$

$$\frac{d}{dt}[Aux_x] = \frac{r_2}{2}[PIN_c][Aux_c] - r_2[Aux_x][PIN_x] - d_{Aux}[Aux_x] \qquad (2)$$

$$\frac{\partial}{\partial t}[CK_c] = D_{ck}\frac{\partial^2}{\partial x^2}[CK] - r_3[Aux_c][CK_c] - d_{CK}[CK_c] \qquad (3)$$

$$\frac{\partial}{\partial t}[CK_x] = D_{ck}\frac{\partial^2}{\partial x^2}[CK] - r_3[Aux_x][CK_x] - d_{CK}[CK_x] \qquad (4)$$

$$\frac{d}{dt}[MP_c] = r_4[Aux_c] - r_9[PXY_a][MP_c] - d_{MP}[MP_c] \qquad (5)$$

$$\frac{d}{dt}[MP_x] = r_4[Aux_x] - d_{MP}[MP_x] \qquad (6)$$

$$\frac{d}{dt}[PIN_c] = r_6[MP_c] - r_5[CK_c][PIN_c] - d_{PIN}[PIN_c] \qquad (7)$$

$$\frac{d}{dt}[PIN_x] = r_6[MP_x] - r_5[CK_x][PIN_x] - d_{PIN}[PIN_x] \qquad (8)$$

$$\frac{d}{dt}[PXY_{in}] = r_7[MP_c] - r_8[PXY_{in}][TDIF] - d_{PXY_{in}}[PXY_{in}] \qquad (9)$$

$$\frac{d}{dt}[PXY_a] = r_8[PXY_{in}][TDIF_p] - d_{PXY_a}[PXY_a] \qquad (10)$$

# 3. Results

For each of the models, with and without MP negative feedback, 7500 parameter sets were chosen at random from the interval (0,20]. A steady state solution was obtained numerically (Methods). The steady state solutions where then compared to the observed concentration profiles of cytokinin and auxin in the root (Figure 1D). As numerical solutions sample small areas of parameter space, closed form, steady state, analysis was also performed. Relationships derived using steady state analysis were verified numerically using the numerical solutions.

## 3.1 Numerical solutions suggest that MP negative feedback improves the ability of the modelled root to achieve a cambial auxin maximum.

As the cytokinin concentration profile did not distinguish between models with and without MP negative feedback, the ability of MP negative feedback to improve the observed auxin maximum in the cambial tissue was tested (Figure 1D). In the model without MP negative feedback 1406 parameter sets (18.75%) had a cambial auxin maximum at steady state, compared to 2172 parameter sets (28.96%) for the model with MP negative feedback. The inclusion of MP negative feedback thus increased the ability of the model to reproduce observed biological data by 10.21% (Figure 2B).

## 3.2 Auxin concentrations in the cambium and phloem determine whether or not the models reproduce the biological data.

To understand how the models which did not reproduce biological data differed from the biological data, steady state concentrations were compared. The cytokinin concentration profile was considered first. For all of the 15000 parameter sets, 7500 for each model, the concentration of cytokinin in the phloem, $[CK_p]$ was greater than the concentration of cytokinin in the cambium, $[CK_c]$ (Figure 2C) and the concentration of cytokinin in the cambium was greater than the concentration of cytokinin in the xylem $[CK_x]$ (Figure 2C). Thus, both models, with and without MP negative feedback, reproduced the cytokinin concentration profile observed *in planta*.

The ability of the models to reproduce the cytokinin concentration profile observed in plants is unsurprising. Cytokinin movement is modelled by diffusion (Hirose et al., 2008). The phloem contains the only source of cytokinin within the models. Cytokinin diffuses from the phloem through the cambium and xylem tissues (Figure 2A). Thus, the only way for cytokinin to be present in the cambium or xylem is via diffusive movement. As such, the concentration of cytokinin in the cambium will be lower than the concentration of cytokinin in the phloem, and the concentration of cytokinin in the xylem will be lower than the concentration of cytokinin in the cambium. There is one reaction effecting the concentrations of cytokinin in the cambium and xylem, the suppression of cytokinin by auxin. Here, auxin negatively regulates cytokinin concentration and signalling (Bishopp et al., 2011a; Mähönen et al., 2006; Moreira et al., 2013; Müller and Sheen, 2008; Nordström et al., 2004; Werner et al., 2006). The suppression of cytokinin by auxin acts to reduce the concentrations of active cytokinin in the cambium and xylem. Therefore, it would be possible for there to be a transient phase in which auxin in the cambium suppressed cambial cytokinin to produce a cytokinin concentration which was lower than the xylem cytokinin concentration. However, as cytokinin in the xylem must be replenished from the cambial cytokinin, via diffusion, over time the xylem cytokinin concentration would drop to below the concentration of cambial cytokinin.

Next the steady state concentrations of auxin were compared. For all 15000 parameter sets the concentration of auxin in the cambium, $[Aux_c]$, was greater than the concentration of auxin in the xylem, $[Aux_x]$ (Figure 2C). Closed form, steady state analysis confirmed that the concentration of auxin in the cambium was always greater than the concentration of auxin in the xylem, for both models (Supporting Information). Thus, the only relationship which determined the ability of the models to reproduce biological data was the comparison of auxin concentrations in the cambium and phloem, $[Aux_p]$ (Figure 2C).

## 3.3 Necessary but not sufficient conditions for cambium auxin concentration to be greater than phloem auxin concentration.

Using closed form, steady state analysis, an inequality was derived which gave a relationship to be satisfied for the auxin concentration in the cambium to be greater than the auxin concentration in the phloem, equation (1)

(Supporting Information). Note that for the model without MP negative feedback $[PXY_a] = r_9 = 0$ giving equation (2) (Figure 1, Methods). Thus, the relationships defining the ability of each model to achieve a cambial auxin maximum differs only by the presence, or not, of MP negative feedback terms, $r_9[PXY_a]$.

$$r_1 > \alpha + \beta \frac{1}{(d_{MP}+r_9[PXY_a])} \tag{11}$$

$$r_1 > \alpha + \beta \frac{1}{(d_{MP})} \tag{12}$$

Where, $d_{Aux} = \alpha$, and $\left(\frac{r_4 r_6 r_2 [Aux_p]}{2(r_5[CK_p]+d_{PIN})}\right) = \beta$.

The relationships defining the ability of each model to achieve a cambial auxin maximum were tested numerically. For the model without MP negative feedback all 1406 parameter sets which had a cambial auxin maximum satisfied equation (12). Similarly, for the model with MP negative feedback, all 2172 parameter sets which had a cambial auxin maximum satisfied equation (11). For both of the models, all parameter sets which did not satisfy either equation (11) or (12), did not have a cambial auxin maximum at steady state.

For the model without MP negative feedback, 766 parameter sets which did not achieve a cambial auxin maximum satisfied equation (12). For the model with MP negative feedback, 731 parameter sets which did not achieve a cambial auxin maximum satisfied equation (11). Thus, the relationships given by equations (11) and (12) are necessary for the models to achieve a cambial auxin maximum (top row, Table 2), but they are not sufficient to guarantee a cambial auxin maximum (first column, Table 2). i.e. if there is a cambial auxin maximum, the parameters will satisfy equations (11) or (12), but equations (11) or (12) cannot be used to generate successful parameter sets.

|  |  | satisfy equation 11 or 12 | |
|---|---|---|---|
|  |  | Y | N |
| cambial auxin maximum | Y | 24% | 0 |
| | N | 10% | 66% |

Table 2: **Analytical relationships necessary but not sufficient.** Percentage of 15000 parameter sets, rounded to two significant figures. Top row: If a parameter set achieves a cambial auxin maximum then it will satisfy ether equations (11) or (12). Bottom row: If a parameter set does not achieve a cambial auxin maximum then it may satisfy ether equations (11) or (12). Left column: If a parameter set satisfies ether equations (11) or (12) it may not achieve a cambial auxin maximum. Right column: If a parameter set does not satisfy ether equations (11) or (12) it will not achieve a cambial auxin maximum.

### 3.4 MP negative feedback always improved the modelled tissues ability to obtain a cambial auxin maximum.

As all parameter sets which achieve a cambial auxin maximum satisfy equations (11) or (12), equations (11) or (12) were used to ask if MP negative feedback always improved the root's ability to obtain a cambial auxin maximum. If MP negative feedback improves the ability of the root model to achieve a cambial auxin maximum, then two things would be expected to be true. One, the addition of MP negative feedback does not prevent a cambial auxin maximum forming. Two, there should be parameter sets for which MP negative feedback is required for the formation of a cambial auxin maximum. Each of these were addressed in turn.

One: Does the addition of MP negative feedback ever prevent a cambial auxin maximum from forming? All parameter values and concentrations are non-negative. Thus, the down-regulation of MP is stronger in the model with MP negative feedback than the model without MP negative feedback,

$$d_{MP} < d_{MP} + r_9[PXY_a] \tag{13}$$

Note, that if the concentration of active PXY, $[PXY_a]$, was equal to zero the model including MP negative feedback would be equal to the model without MP negative feedback. Thus, for this analysis, concentrations of active PXY greater than zero were considered. It follows that,

$$\frac{1}{d_{MP}} > \frac{1}{(d_{MP}+r_9[PXY_a])} \tag{14}$$

Because all parameter values and concentrations are non-negative $\alpha$ and $\beta$ are also non-negative. Considering nontrivial values of $\alpha$ and $\beta$, which are greater than zero, gives the relationship,

$$\alpha + \beta \frac{1}{d_{MP}} > \alpha + \beta \frac{1}{(d_{MP}+r_9[PXY_a])} \tag{15}$$

The left-hand side of equation (15) is equal to the right-hand side of equation (12). Thus, if the model without MP negative feedback achieves a cambial auxin maximum, the parameters of that model will satisfy equation (12) and equations (15) and (12) can be combined to get equation (16).

$$r_1 > \alpha + \beta \frac{1}{d_{MP}} > \alpha + \beta \frac{1}{(d_{MP}+r_9[PXY_a])} \tag{16}$$

Equation (16) states that if the model without MP negative feedback achieves a cambial auxin maximum, then adding MP negative feedback to that model will satisfy both equation (11), and have a cambial auxin maximum.

The analysis states that, MP negative feedback does not prevent a cambial auxin maximum forming. This analysis was tested numerically. For model without MP negative feedback, 1406 parameter sets had a cambial auxin maximum at steady state (Figure 2B). MP negative feedback was added to these 1406 parameter sets by choosing values, greater than zero, for MP negative feedback parameters ($[TDIF_p], r_7, r_8, r_9, d_{PXY_{in}}, d_{PXY_a}$). Ten sets of MP negative feedback parameters were chosen uniformly at random from the interval (0,20], for each of the 1406 parameter sets, resulting in 14060 parameter sets. The model with MP negative feedback was solved for each of the 14060 parameter sets. All 14060 parameter sets achieved a cambial auxin maximum at steady state confirming the analytical results.

Two: Are there parameter sets for which MP negative feedback is required for the formation of a cambial auxin maximum? As there are no parameter sets which do not satisfy equations (11) or (12) and achieve a cambial auxin maximum (Table 2), parameter sets which do not satisfy equation (12) will be considered in the following analysis. For MP negative feedback to be required there must be a set of parameters for which the model without MP negative feedback does not achieve a cambial auxin maximum but the model with MP negative feedback does. For the model without MP negative feedback, parameter sets which do not satisfy equation (12) do not obtain a cambial auxin maximum. Those parameters satisfy equation (17),

$$\alpha + \beta \frac{1}{(d_{MP})} \geq r_1 \tag{17}$$

For the model with MP negative feedback to obtain a cambial auxin maximum the parameters must satisfy equation (11). Merge equations (17) and (11) to get,

$$\alpha + \beta \frac{1}{(d_{MP})} \geq r_1 > \alpha + \beta \frac{1}{(d_{MP}+r_9[PXY_a])} \tag{18}$$

Equation (18) does not contradict equation (15), which is always true. Equation (18) states that there are values of $r_9$ and $[PXY_a]$ for which the model with MP negative feedback will obtain a cambial auxin maximum, but the cambial auxin maximum would be lost if the MP negative feedback loop were removed. i.e. there are a set of parameters for which MP negative feedback is required to obtain a cambial auxin maximum. This analysis was tested numerically. There were 2172 parameter sets for the model with MP negative feedback which enabled the model to achieve a cambial auxin maximum. For each of those parameter sets the MP negative feedback parameters ($[TDIF_p], r_7, r_8, r_9, d_{PXY_{in}}, d_{PXY_a}$) were set to zero. The model without MP negative feedback was then solved using the 2172 adjusted parameter sets. 801 (36.88%) of the adjusted parameter sets did not achieve a cambial auxin maximum. Thus, 10.68% of the 7500 parameter sets chosen at random, from the interval (0,20], were required to achieve a cambial auxin maximum in the model with MP negative feedback.

### 3.5 If MP is stable, MP negative feedback becomes a requirement for the modelled root to form a cambial auxin maximum.

The left and right terms of Equation (18) define an interval, $I$, which contains parameters sets for which the loop was required. It was hypothesised that an increase, or decrease, in the size of interval $I$ would increase, or decrease, the number of parameters sets for which the loop was required. To find the parameters which had an effect on the size of the interval $I$ an equation for the length of the interval was defined (equation 19) and rearranged (equation 20).

$$I = \alpha + \beta \frac{1}{d_{MP}} - \left(\alpha + \beta \frac{1}{(d_{MP}+r_9 PXY_a)}\right) \tag{19}$$

$$I = \beta \frac{1}{d_{MP}} \frac{r_9 PXY_a}{(d_{MP} + r_9 PXY_a)} \tag{20}$$

Equation (20) states that; the length of interval $I$ approaches infinity as the basal degradation of MP, $d_{MP}$, approaches zero; the length of $I$ approaches zero as $d_{MP}$ approaches infinity. The analysis was tested numerically for different intervals of MP basal degradation, $d_{MP}$. The intervals were, (0,1000], (0,100], (0,10], (0,1] and (0,0.1]. For each of the $d_{MP}$ intervals, for each of the models, with and without MP negative feedback, 7500 parameter sets were chosen at random. All parameters, apart from $d_{MP}$, were chosen from the interval (0,20], as before. $d_{MP}$ was chosen from the specified $d_{MP}$ intervals (0,1000], (0,100], (0,10], (0,1] and (0,0.1]. A steady state solution was obtained numerically for each parameter set (Methods). The numerical results confirmed the analysis, the percentage of required parameter sets for which MP negative feedback was required increased as $d_{MP}$ was chosen from an interval closer to 0, i.e. as the size of interval $I$ was increased (Figure 3, Table 3). Thus, as the basal degradation of MP is reduced, MP negative feedback becomes a stronger requirement for the modelled tissue to achieve a cambial auxin maximum.

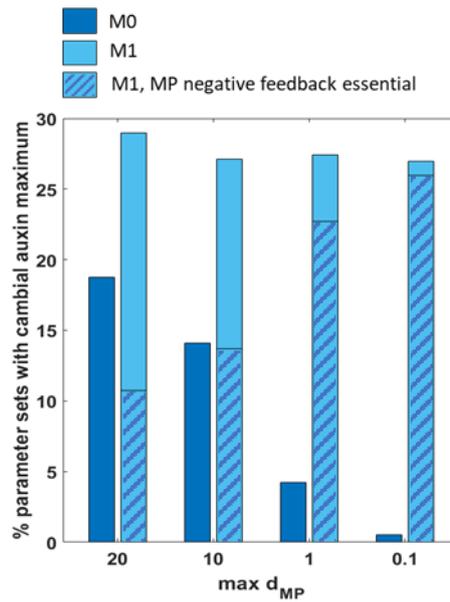

**Figure 3. MP negative feedback becomes essential if MP is stable.** Percentage of parameter sets that achieve cambium auxin maximum in the presence (M1) or absence (M0) of MP negative feedback. Parameter space for MP degradation is increased by an order of magnitude for each of the 7500 parameter sets. All other parameters were chosen from the interval (0,20]. The hashed bars show the percentage of parameter sets for which MP negative feedback was essential for a cambial auxin maxima to be achieved.

|  | $d_{MP}$ | | | | |
| --- | --- | --- | --- | --- | --- |
|  | (0, 0.1] | (0, 1] | (0, 10] | (0, 100] | (0, 1000] |
| **M0** | 40 (0.53%) | 318 (4.24%) | 1056 (14.08%) | 2327 (31.03%) | 3380 (45.07%) |
| **M1** | 2023 (26.97%) | 2058 (27.44%) | 2033 (27.11%) | 2538 (33.84%) | 3321 (44.28%) |
| **M1 essential** | 1948 (25.97%) | 1703 (22.71%) | 1027 (13.69%) | 311 (4.15%) | 38 (0.51%) |

**Table 3. MP negative feedback becomes essential if MP is stable.** Percentage of parameter sets that achieve cambium auxin maximum in the presence (M1) or absence (M0) of MP negative feedback. Parameter space for MP degradation is increased by an order of magnitude for each of the 7500 parameter sets. All other parameters were chosen from the interval (0,20].

## 4. Discussion

### 4.1 Modelled interactions can replicate hormone patterns in the cambium.

Auxin was first proposed to influence vascular development in the 50's via exogenous applications (Torrey, 1953), and was subsequently found to occur in a concentration maximum in the cambium of pine trees (Uggla et al., 1996). This observation led to the suggestion that auxin can act as a positional signal in plant vascular tissue, and the pattern of auxin accumulation was found to hold true for a number of vascular plant species including (Brackmann et al., 2018). At sites of auxin accumulation within plant tissues, responses are mediated by the ARF family of transcription factors. In Arabidopsis, MP is a key member of this family (Roosjen et al., 2018). Recent studies in Arabidopsis have dissected the role of MP in the cambium, both early in development during cambium formation in seedlings, and later in development to maintain the established cambium. Surprisingly, MP was reported to have opposing functions at these two stages of development. During cambium initiation in seedlings, MP promoted cambial divisions via activation of the PXY receptor kinase (Smetana et al., 2019). In established tissue, MP repressed cell division in the cambium, and PXY was found to repress MP activation (Brackmann et al., 2018; Han et al., 2018).

One explanation for both MP activities (promotion and repression of cambial cell division) is that MP and PXY form a negative feedback loop (Figure 1C). A mathematical model was generated to understand if MP negative feedback had consequences for the robustness of the system. The model was characterised by reactions in three domains, phloem, cambium, and xylem, with each domain sized to match measured cell sizes in Arabidopsis hypocotyls. The consequences of MP negative feedback were gauged by determining the number of parameter sets in which auxin was correctly patterned, with an auxin maxima in the cambium, in the presence of MP negative feedback, compared to its absence. The presence of MP negative feedback resulted in a higher proportion of parameter sets generating an auxin maximum in the cambium, 28.96% relative to 18.75% without (Figure 2B). As such, inclusion of MP negative feedback increased the ability of the model to reproduce the auxin maxima first observed in the pine cambium 25 years ago. Further support for the MP negative feedback loop came from two sets of further analysis of those parameters for which an auxin cambial maximum was reached. Firstly, for parameter sets that generated the cambial auxin maxima in the absence of MP negative feedback, subsequent addition of the feedback loop never disrupted the system such that the auxin maxima was lost. Thus, MP negative feedback did not impede auxin maximum formation in the model. Secondly, for those parameter sets that generated the cambial auxin maximum in the presence of MP negative feedback, for 10.68% of those tested, removal of the MP feedback loop resulted in loss of the cambium auxin maximum. As such, the theoretical evidence described here, supports the notion that the MP negative feedback loop increases the ability of the system to pattern correctly.

## 4.2 Auxin acts with receptor kinases to balance stem cell fate

The PXY signalling mechanism is closely related to others that regulate stem cell fate in plant meristems. Among the best characterised of these is the CLAVATA (CLV) system, which regulates the shoot apical meristem. Here CLV3 ligand (related to TDIF), signals to CLV1 (related to PXY) and regulates expression of *WUSCHEL* (*WUS*), a homeodomain transcription factor (Brand et al., 2000; Fletcher et al., 1999; Ogawa et al., 2008). ChIP-seq data suggests that WUS directly controls expression of components of auxin transport and response. Auxin is essential at low levels to maintain the shoot apical meristem stems cells (Ma and Li, 2019). However, its presence at higher levels is also necessary for organ formation and the flanks of the stem cells in the shoot meristem (Reinhardt et al., 2000; Vernoux et al., 2000). Thus, the shoot apical meristem represents another system in which the auxin response must be balanced to facilitate differing outcomes, and in this stem cell population, peptide-receptor signalling appears to contribute to balancing these outcomes (Ma and Li, 2019). The relationship between auxin and peptide-receptor signalling is likely to be ancient, as in the moss *Physcomitrium*, stem cells are regulated PpRPK2, a PXY/CLV relative, that regulates auxin homeostasis (Nemec-Venza et al., 2022).

## 4.3 Future directions

A striking feature of homeostatic mechanisms in plant meristems, both vascular and apical, is that they must be effective across developmental time. In the case of the cambium, that also must include efficacy across scales. The cambium initiates as cell divisions in xylem adjacent cells, but it can increase in size to multiple cells within a single cell file. While some such variability is present in Arabidopsis, in species such as poplar, in which TDIF-PXY and auxin have both been shown to act in the cambium (Etchells et al., 2015; Kucukoglu et al., 2017; Nilsson et al., 2008; Xu et al., 2019), the number of cambium cells fluctuates further, with rapid cambial expansion occurring in early summer prior to cessation of division during winter. Thus, manipulation of the feedback mechanisms proposed here, possibly by other factors, may influence cambium cell number. These factors could include alterations to feedback caused by differences in tissue topology. For example, high numbers of cambium cells in a cell file (as is seen in established, actively growing cambium) would likely induce changes to cytokinin and auxin gradients due to greater physical distances between the source of hormone and at least a subset of the dividing cells in the cambium. The work described here thus acts as a framework to test how the outcomes might differ in two-dimensional space in morphologically accurate tissues at differing stages of cambium development. Such future work also represents an opportunity to refine aspects of the model. For example, one limitation includes the well-mixed assumption. The summative PIN lacking positioning restricts the ability to observe the details of auxin transport under the influence of PXY and MP interactions.

Regarding the perceived change in MP activity throughout development, we hypothesise that other molecular factors have an effect on MP activation of PXY or TDIF-PXY repression of MP to control cambium size during development. It remains an open question for further research, what these factors might be. The requirement for MP negative feedback was high in parameter sets in which basal degradation of MP was reduced (Figure 3). As such, it might also be interesting to determine MP turnover rates in the vascular cambium at different developmental stages.

## Supplementary Information

## Steady state analysis

Steady state analysis was used to generate parameter relationships for which a cambial auxin maximum can be achieved within the model containing MP negative feedback, referred to as model M1, (equations (1) to (10) in the main manuscript). These M1 parameter relationships were then used to generate parameter relationships for a cambial auxin maximum in the model without MP negative feedback, model M0, (equations (1) to (10) in the main manuscript with $[TDIF_p], r_7, r_8, r_9, d_{PXY_{in}}, d_{PXY_a}$ set equal to zero). The inequalities $[Aux_c] > [Aux_p]$ and $[Aux_c] > [Aux_x]$ must be satisfied for there to be a cambial auxin maximum.

As discussed in the manuscript, the modelled cytokinin concentration profile always has a peak in the phloem and thus satisfies the inequality $[CK_p] > [CK_c] > [CK_x]$. As such the inequality $[CK_p] > [CK_c] > [CK_x]$ was used in the steady state analysis rather than the partial differential equations describing $CK_c$ and $CK_p$ concentrations (equations (3) and (4) in the main manuscript).

Note that all concentrations and the six parameters involved in monopteros negative feedback ($TDIF_p, d_{PXY_a}, d_{PXY_{in}}, r_7, r_8, r_9$) are greater than or equal to zero. The remaining twelve parameters are greater than zero ($[Aux_p], [CK_p], r_1, r_2, r_3, r_4, r_5, r_6, d_{Aux}, d_{CK}, d_{PIN}, d_{MP}$).

At steady state, equations (1), (2), (5), (6), (7), (8), (9) and (10) become,

$$r_1[Aux_p] + \frac{r_2}{2}[PIN_x][Aux_x] - r_2[Aux_c][PIN_c] - d_{Aux}[Aux_c] \tag{S1}$$

$$\frac{r_2}{2}[PIN_c][Aux_c] - r_2[Aux_x][PIN_x] - d_{Aux}[Aux_x] \tag{S2}$$

$$r_4[Aux_c] - r_9[PXY_a][MP_c] - d_{MP}[MP_c] \tag{S3}$$

$$r_4[Aux_x] - d_{MP}[MP_x] \tag{S4}$$

$$r_6[MP_c] - r_5[CK_c][PIN_c] - d_{PIN}[PIN_c] \tag{S5}$$

$$[MP_x] - r_5[CK_x][PIN_x] - d_{PIN}[PIN_x] \tag{S6}$$

$$r_7[MP_c] - r_8[PXY_{in}][TDIF] - d_{PXY_{in}}[PXY_{in}] \tag{S7}$$

$$r_8[PXY_{in}][TDIF_p] - d_{PXY_a}[PXY_a] \tag{S8}$$

### Auxin in the cambium is always greater than auxin in the xylem.

Here, it is shown that the inequality $Aux_c > Aux_x$ always holds at steady state. This is achieved by showing that the contradiction inequality $Aux_c \leq Aux_x$ results in two opposing relationships and thus $Aux_c \leq Aux_x$ cannot be true.

### An overview of the analysis to follow.

Assume that the contradiction inequality (S9) holds.

$$[Aux_c] \leq [Aux_x] \tag{S9}$$

Step 1: Equations (S3), (S4) and (S9) are used to derive the inequality $[MP_c] \leq [MP_x]$, (S15). (S15) is true for models M1 and M0 and is a direct result of the contradiction inequality (S9). Step 2: Equations (S5), (S6), (S15) and the inequality $[CK_c] > [CK_x]$ are used to derive the inequality $PIN_c < PIN_x$, (S22). Step 3: Equations (S1), (S2) and (S9) are used to derive the inequality $PIN_c > PIN_x$, (S29).

The inequalities $PIN_c < PIN_x$ (S22) and $PIN_c > PIN_x$ (S29) contradict each other, indicating that $[Aux_c] \leq [Aux_x]$ and $[CK_c] > [CK_x]$ cannot both hold. As $[CK_c] > [CK_x]$ is true, $[Aux_c] \leq [Aux_x]$ must be false. Thus, the concentration of auxin in the cambium is greater than the concentration of auxin in the xylem. This result holds for both models M1 and M0.

The analysis.

### Step 1.

Equations (S3), (S4) and (S9) are used to derive the inequality $[MP_c] \leq [MP_x]$, (S15).

Rearrange equations (S3) and (S4) to obtain expressions for $[Aux_c]$ and $[Aux_x]$.

$$[Aux_c] = \frac{1}{r_4}[MP_c](r_9[PXY_a] + d_{MP}) \tag{S10}$$

$$[Aux_x] = \frac{1}{r_4}[MP_x]d_{MP} \tag{S11}$$

Substitute (S10) and (S11) into (S9) and multiply by $r_4 > 0$.

$$[MP_c](r_9[PXY_a] + d_{MP}) \leq [MP_x]d_{MP} \tag{S12}$$

Divide (S12) by $d_{MP} > 0$.

$$[MP_c]\left(\frac{r_9[PXY_a]}{d_{MP}} + 1\right) \leq [MP_x] \tag{S13}$$

$\frac{r_9[PXY_a]}{d_{MP}} \geq 0$ for model M1 and $\frac{r_9[PXY_a]}{d_{MP}} = 0$ for model M0. Thus, $\left(\frac{r_9[PXY_a]}{d_{MP}} + 1\right) \geq 1$ and,

$$[MP_c] \leq [MP_c]\left(\frac{r_9[PXY_a]}{d_{MP}} + 1\right) \leq [MP_x] \tag{S14}$$

giving,

$$[MP_c] \leq [MP_x] \tag{S15}$$

Note that for M0, $r_9[PXY_a] = 0$, and $[MP_c] \leq [MP_x]$ is obtained from equation (S13). The inequality (S15) is a direct result of the contradicting inequality, (S9).

### Step 2.

Equations (S5), (S6), (S15) and the inequality $[CK_c] > [CK_x]$ are used to derive the inequality $PIN_c < PIN_x$, (S22).

Rearrange (S5) and (S6).

$$[MP_c] = \frac{1}{r_6}[PIN_c](r_5[CK_c] + d_{PIN}) \tag{S16}$$

$$[MP_x] = \frac{1}{r_6}[PIN_x](r_5[CK_x] + d_{PIN}) \tag{S17}$$

Substitute (S16) and (S17) into (S15) and rearrange.

$$\frac{1}{r_6}[PIN_c](r_5[CK_c] + d_{PIN}) - \frac{1}{r_6}[PIN_x](r_5[CK_x] + d_{PIN}) \leq 0 \tag{S18}$$

As $[CK_c] > [CK_x]$,

$$\frac{1}{r_6}[PIN_c](r_5[CK_x] + d_{PIN}) < \frac{1}{r_6}[PIN_c](r_5[CK_c] + d_{PIN}) \tag{S19}$$

Merge (S18) and (S19) to get,

$$\frac{1}{r_6}[PIN_c](r_5[CK_x] + d_{PIN}) - \frac{1}{r_6}[PIN_x](r_5[CK_x] + d_{PIN}) < 0 \tag{S20}$$

Multiply (S20) by $r_6$ and rearrange.

$$([PIN_c] - [PIN_x])(r_5[CK_x] + d_{PIN}) < 0 \tag{S21}$$

$(r_5[CK_x] + d_{PIN}) > 0$, thus, $[PIN_c] - [PIN_x] < 0$, giving,

$$[PIN_c] < [PIN_x] \tag{S22}$$

### Step 3.

Equations (S1), (S2) and (S9) are used to derive the inequality $PIN_c > PIN_x$, (S29).

Considering the degradation term in (S1), use (S9) to obtain the inequality,

$$-d_{Aux}[Aux_c] \geq -d_{Aux}[Aux_x] \tag{S23}$$

Merge (S1) and (S23).

$$0 \geq r_1[Aux_p] + \frac{r_2}{2}[PIN_x][Aux_x] - r_2[Aux_c][PIN_c] - d_{Aux}[Aux_x] \tag{S24}$$

Obtain an expression containing $[PIN_x]$ and $[PIN_c]$ by rearranging (S2) to get an expression for $d_{Aux}[Aux_x]$.

$$d_{Aux}[Aux_x] = \frac{r_2}{2}[PIN_c][Aux_c] - r_2[Aux_x][PIN_x] \tag{S25}$$

Substitute (S25) into (S24) and rearrange.

$$0 \geq r_1[Aux_p] + \frac{3}{2}r_2([PIN_x][Aux_x] - [Aux_c][PIN_c]) \tag{S26}$$

Using (S9) derive the inequality,

$$[PIN_x][Aux_x] \geq [PIN_x][Aux_c] \tag{S27}$$

Merge (S26) and (S27), then rearrange.

$$0 \geq r_1[Aux_p] + \frac{3}{2}r_2[Aux_c]([PIN_x] - [PIN_c]) \tag{S28}$$

Recall, $r_1, [Aux_p], r_2 > 0$. Thus, $r_1[Aux_p] > 0$. $\frac{3}{2}r_2[Aux_c] = 0$ if $[Aux_c] = 0$, $\frac{3}{2}r_2[Aux_c] > 0$ otherwise. As $r_1[Aux_p] > 0$, $[Aux_c] > 0$ for (S28) to be satisfied. Thus, for inequality (S28) to hold $[Aux_c] > 0$ and $([PIN_x] - [PIN_c]) < 0$. i.e.

$$[PIN_x] < [PIN_c] \tag{S29}$$

(S22) and (S29) contradict each other, indicating that $[Aux_c] \leq [Aux_x]$ and $[CK_c] > [CK_x]$ cannot both hold. As $[CK_c] > [CK_x]$ is true, $[Aux_c] \leq [Aux_x]$ must be false. Thus, the concentration of auxin in the cambium is greater than the concentration of auxin in the xylem. This result holds for both models M1 and M0.

## Conditions for auxin in the cambium to be greater than auxin in the phloem.

Here, an inequality, (S42), is derived that describes a relationship which is satisfied when auxin in the cambium is greater than auxin in the phloem, $[Aux_c] > [Aux_p]$. (S42) is different for M1 and M0 and thus can be used to compare the two models.

### An overview of the analysis to follow.

(S1) is the only equation containing both $[Aux_c]$ and $[Aux_p]$. Thus, in order to derive a condition for which the inequality $[Aux_c] > [Aux_p]$ holds, (S1) must be used. Step 1: Rearrange (S2) to obtain an expression for $[Aux_x]$. Substitute $[Aux_x]$ out of (S1), then rearrange to derive the inequality (S32). Step 2: Use the auxin condition to be satisfied, $[Aux_c] > [Aux_p]$, to remove auxin from (S32), resulting in the inequality (S35). Step 3: The inequality (S35) does not distinguish between models M1 and M0. Equations (S5), (S3) and the inequalities $[CK_p] > [CK_c]$, $[Aux_c] > [Aux_p]$ are used to generate an inequality, (S42), which can distinguish between M1 and M0 and thus can be used to compare the two models. Step 4: The conversion of (S42) into equation (11) in the main manuscript.

### The analysis.

*Step 1.*

Rearrange (S2) and substitute $[Aux_x]$ out of (S1), then rearrange to derive the inequality (S32).
$d_{Aux} > 0$, $Aux_x \geq 0$, thus (S2) can be used to obtain the inequality,

$$\frac{r_2}{2}[Aux_c][PIN_c] - r_2[Aux_x][PIN_x] \geq 0 \tag{S29}$$

Rearrange (S29) and divide by $r_2 > 0$, to get,

$$[Aux_c][PIN_c] \geq 2[Aux_x][PIN_x] \tag{S30}$$

giving,

$$[Aux_c][PIN_c] \geq [Aux_x][PIN_x] \tag{S31}$$

Substitute (S31) into (S1) and rearrange to get,

$$r_1[Aux_p] - [Aux_c]\left(\frac{r_2}{2}[PIN_c] + d_{Aux}\right) \geq 0 \tag{S32}$$

*Step 2.*

Step 2: Use the auxin condition to be satisfied, $[Aux_c] > [Aux_p]$, to remove auxin from (S32), resulting in the inequality (S35).

As $[Aux_c] > [Aux_p]$ and $r_1 > 0$,

$$r_1[Aux_c] > r_1[Aux_p] \tag{S33}$$

Substitute (S33) into (S32) to get,

$$r_1[Aux_c] - [Aux_c]\left(\frac{r_2}{2}[PIN_c] + d_{Aux}\right) > 0 \tag{S34}$$

As $[Aux_p] > 0$ and parameters which satisfy the condition $[Aux_c] > [Aux_p]$ are of interest, $[Aux_c] > 0$. Divide (S34) by $[Aux_c]$.

$$r_1 - \left(\frac{r_2}{2}[PIN_c] + d_{Aux}\right) > 0 \tag{S35}$$

*Step 3.*

The inequality (S35) does not distinguish between models M1 and M0. Equations (S5), (S3) and the inequalities $[CK_p] > [CK_c]$, $[Aux_c] > [Aux_p]$ are used to generate an inequality, (S42), which can distinguish between M1 and M0 and thus can be used to compare the two models.

Rearrange (S5) to give an expression for $[PIN_c]$.

$$[PIN_c] = \frac{r_6[MP_c]}{(r_5[CK_c] + d_{PIN})} \tag{S36}$$

Substitute (S36) into (S35).

$$r_1 - \left(\frac{r_6 r_2[MP_c]}{2(r_5[CK_c] + d_{PIN})} + d_{Aux}\right) > 0 \tag{S37}$$

Use the inequality $[CK_p] > [CK_c]$ to replace the steady state concentration, $[CK_c]$, with the parameter, $[CK_p] > 0$. Note that as $r_6, r_5, r_2, [MP_c], [CK_p], d_{PIN} \geq 0$,

$$\frac{r_6 r_2[MP_c]}{2(r_5[CK_p] + d_{PIN})} \geq \frac{r_6 r_2[MP_c]}{2(r_5[CK_c] + d_{PIN})} \tag{S38}$$

Substitute (S38) into (S37).

$$r_1 - \left(\frac{r_6 r_2[MP_c]}{2(r_5[CK_p] + d_{PIN})} + d_{Aux}\right) > 0 \tag{S39}$$

Equation (S3) contains the components of model M1 which are responsible for MP self-repression in the cambium. Thus, rearranging (S3) to obtain an expression for $[MP_c]$, then substituting this expression into (S39) will bring MP self-repression into the resulting inequality.

Rearrange (S3) to get an expression for $[MP_c]$.

$$[MP_c] = \frac{r_4[Aux_c]}{r_9[PXY_a] + d_{MP}} \tag{S40}$$

Substitute (S40) into (S39) and rearrange.

$$r_1 > d_{Aux} + \frac{r_4 r_6 r_2[Aux_c]}{2(r_5[CK_p] + d_{PIN})}\left(\frac{1}{r_9[PXY_a] + d_{MP}}\right) \tag{S41}$$

$[Aux_c]$ is a final steady state concentration. A relationship satisfying the condition $[Aux_c] > [Aux_p]$ is of interest. Thus, the condition $[Aux_c] > [Aux_p]$ is used to substitute $[Aux_c]$ out of (S41).

$$r_1 > d_{Aux} + \frac{r_4 r_6 r_2[Aux_p]}{2(r_5[CK_p] + d_{PIN})}\left(\frac{1}{r_9[PXY_a] + d_{MP}}\right) \tag{S42}$$

$[PXY_a]$ is the only steady state concentration remaining in (S42). An inequality for $[PXY_a]$ in terms of parameters was derived (not shown here). However, the inequality was complex and did not give any further insight than (S42). Thus, (S42) was used to compare the two models.

*Step 4.*

The conversion of (S42) into (11), from the main manuscript.

Let, $d_{Aux} = \alpha$, and $\left(\frac{r_4 r_6 r_2 [Aux_p]}{2(r_5[CK_p]+d_{PIN})}\right) = \beta$. (S42) becomes,

$$r_1 > \alpha + \beta \frac{1}{(d_{MP}+r_9 PXY_a)} \tag{11}$$

# Numerical Scheme for diffusion equation

The model contained three, well-mixed, spatial domains; the phloem, cambium and xylem. The lengths of the domains were set to; $2.6 \mu m$ for the phloem, $1.2 \mu m$ for the cambium and $10.8 \mu m$ for the xylem (See section 2.1 Model formulation) (Bagdassarian et al., 2020; Wang et al., 2019). The hormone cytokinin was the only component of the model which moved between the domains via diffusion. Cytokinin is rapidly transported down the phloem from the shoot to the root (Hirose et al., 2008) and is thus modelled as a constant cytokinin sauce, $[CK_p]$. Cytokinin then diffuses from the phloem into the cambium, $[CK_c]$, and xylem, $[CK_x]$. The diffusion coefficient of cytokinin was set to $D_{CK} = 220 \; \mu m^2/s$ (Moore et al., 2015). Any cytokinin moving from the xylem into the central tissues is removed from the model. Thus, the diffusion equation was solved on an irregular mesh with four mesh points (Figure S1). The first three mesh points represent the centre of each of the three domains, while the final mesh point represents the central tissues on the opposite side of the xylem to the cambium, the central tissues. The cytokinin concentration in the central tissues was not calculated as it was removed from the model. Furthermore, no cytokinin moved into the xylem from the central tissues. $\Delta x_1 = 1.9 \mu m$ is the distance from the centre of the phloem to the centre of the cambium. $\Delta x_1 = 6 \mu m$ is the distance from the centre of the cambium to the centre of the xylem. $\Delta x_3 = 5.4 \; \mu m$ is the distance from the centre of the xylem to the central tissues. Let $\Delta t$ represent the iterative time step for the finite difference method and $t = 0,1,2,3, \ldots$ be the time index.

Cytokinin in the phloem is constant,
$$[CK_p^{t+1}] = [CK_p^t] \tag{S43}$$

The finite difference scheme for the effect of diffusion on the concentration of cytokinin in the cambium is,
$$[CK_c^{t+1}] = [CK_c^t] - D_{ck}\Delta t \left(\frac{1}{(\Delta x_1)^2}[CK_c^t] + \frac{1}{(\Delta x_2)^2}[CK_c^t]\right) + D_{ck}\frac{\Delta t}{(\Delta x_1)^2}[CK_p^t] + D_{ck}\frac{\Delta t}{(\Delta x_2)^2}[CK_x^t] \tag{S44}$$

The finite difference scheme for the effect of diffusion on the concentration of cytokinin in the xylem is,
$$[CK_x^{t+1}] = [CK_x^t] + -D_{ck}\Delta t \left(\frac{1}{(\Delta x_2)^2}[CK_x^t] + \frac{1}{(\Delta x_3)^2}[CK_x^t]\right) + D_{ck}\frac{\Delta t}{(\Delta x_2)^2}[CK_c^t] \tag{S45}$$

Equations (S43) to (S45) can be brought together to form the matrix equation,

$$\begin{bmatrix}[CK_p^{t+1}]\\[CK_c^{t+1}]\\[CK_x^{t+1}]\end{bmatrix} = \begin{bmatrix} 1 & 0 & 0 \\ \left(D_{ck}\frac{\Delta t}{(\Delta x_1)^2}\right) & \left(1 - D_{ck}\Delta t\left(\frac{1}{(\Delta x_1)^2} + \frac{1}{(\Delta x_2)^2}\right)\right) & \left(D_{ck}\frac{\Delta t}{(\Delta x_2)^2}\right) \\ 0 & \left(D_{ck}\frac{\Delta t}{(\Delta x_2)^2}\right) & \left(1 - D_{ck}\Delta t\left(\frac{1}{(\Delta x_2)^2} + \frac{1}{(\Delta x_3)^2}\right)\right) \end{bmatrix} \begin{bmatrix}[CK_p^t]\\[CK_c^t]\\[CK_x^t]\end{bmatrix} \tag{S36}$$

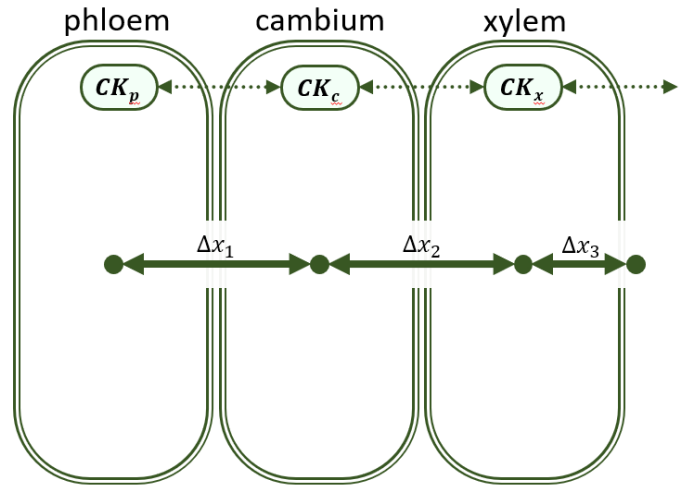

***Figure S1: The irregular mesh used to solve the diffusion equation.*** *The first three mesh points (from left to right) represent the centre of each of the three domains, while the final mesh point represents the central tissues on the opposite side of the xylem to the cambium, the central tissues. $\Delta x_1 = 1.9 \mu m$ is the distance from the centre of the phloem to the centre of the cambium. $\Delta x_1 = 6 \mu m$ is the distance from the centre of the cambium to the centre of the xylem. $\Delta x_3 = 5.4\ \mu m$ is the distance from the centre of the xylem to the central tissues.*